\documentclass[useAMS,usenatbib,usegraphicx]{mn2e}

\usepackage{epsfig}

\title[The local FIR Galaxy Colour-Luminosity Distribution]{The local
  FIR Galaxy Colour-Luminosity distribution: A reference for BLAST,
  and {\it Herschel}/SPIRE sub-mm surveys}

\author[E.~L. Chapin, D.~H. Hughes, I. Aretxaga] 
{E.L.~Chapin$^{1}$, D.H.~Hughes$^{2}$, I.~Aretxaga$^{2}$ \\
  $^{1}$Dept. of Physics \& Astronomy, Univ. of British Columbia,
  6224 Agricultural Road, Vancouver, B.C. V6T 1Z1, Canada\\
  $^{2}$Instituto Nacional de Astrof\'isica, \'Optica y Electr\'onica (INAOE),
  Aptdo. Postal 51 y 216, Puebla, Mexico\\
}

\def\lsim{\mathrel{\lower2.5pt\vbox{\lineskip=0pt\baselineskip=0pt
           \hbox{$<$}\hbox{$\sim$}}}}

\def\gsim{\mathrel{\lower2.5pt\vbox{\lineskip=0pt\baselineskip=0pt
           \hbox{$>$}\hbox{$\sim$}}}}

\begin{document}

\label{firstpage}

\maketitle

\begin{abstract}

  We measure the local galaxy far-infrared (FIR) 60-to-100\,\micron\
  colour-luminosity distribution using an all-sky {\it IRAS}
  survey. This distribution is an important reference for the next
  generation of FIR--submillimetre surveys that have and will conduct
  deep extra-galactic surveys at 250--500\,\micron. With the peak in
  dust-obscured star-forming activity leading to present-day giant
  ellipticals now believed to occur in sub-mm galaxies near
  $z\sim2.5$, these new FIR--submillimetre surveys will directly
  sample the SEDs of these distant objects at rest-frame FIR
  wavelengths similar to those at which local galaxies were observed
  by {\it IRAS}. We have taken care to correct for temperature bias
  and evolution effects in our {\it IRAS} 60\,\micron-selected
  sample. We verify that our colour-luminosity distribution is
  consistent with measurements of the local FIR luminosity function,
  before applying it to the higher-redshift Universe. We compare our
  colour-luminosity correlation with recent dust-temperature
  measurements of sub-mm galaxies and find evidence for pure
  luminosity evolution of the form $(1+z)^3$. This distribution will
  be useful for the development of evolutionary models for BLAST and
  SPIRE surveys as it provides a statistical distribution of
  rest-frame dust temperatures for galaxies as a function of
  luminosity.
 
\end{abstract}

\begin{keywords}
  galaxies: luminosity function, infrared: galaxies, submillimetre,
  galaxies: evolution
\end{keywords}

\section{Introduction}

Deep extra-galactic surveys at sub-mm wavelengths
($\sim200$--1200\,\micron) over the last 10 years have uncovered a
population of luminous infrared galaxies ($L>10^{12}$\,$L_\odot$) with
star-formation rates inferred to be $\rm \gg 1000 M_{\odot}/yr^{-1}$
\citep[e.g.][]{smail1997,hughes1998,barger1998,eales1999,cowie2002,scott2002,borys2003,serjeant2003,webb2003,wang2004,greve2004,laurent2005,coppin2005,coppin2006,knudsen2006,bertoldi2007,scott2008,greve2008,perera2008}. These
sub-mm galaxies (SMGs hereafter) are believed to be high-redshift
($z>1$) analogues, and in many cases more luminous examples, of local
Ultra-Luminous Infrared Galaxies (ULIRGs) discovered with {\it IRAS}
20 years ago \citep{sanders1996}. Furthermore, the rest-frame
wavelengths sampled by sub-mm surveys of the highest-redshift SMGs
approaches those of the far-infrared (FIR) {\it IRAS} observations.
Appearing in vast quantities consistent with massive evolution of the
local ULIRG population, these SMGs are now believed to represent an
important early stage in the evolutionary sequence that ultimately
produces locally-observed massive elliptical galaxies
\citep[e.g.][]{scott2002,blain2004}. Thus, to this day, the local
far-infrared luminosity function measured by {\it IRAS} continues to
be useful for interpreting the results of these longer-wavelength
surveys.

Additional motivation for studying the {\it IRAS} luminosity function,
and its connection to the higher-redshift SMG population, comes from
the shape and magnitude of the cosmic infrared background (CIB)
measured by COBE which peaks near 200\,\micron.
\citep[e.g.][]{fixsen1998}. Its broad shape resembles the
superposition of many thermal SEDs, which can be interpreted as
evidence for a population of sources at redshifts $z<1$ with a large
range of physical temperatures, or alternatively, as a population with
a narrower range of temperatures, but residing over a greater range of
redshifts, including a significant fraction at $z>1$ (the SMG
population). This latter possibility is supported by the fact that the
total energy density of the CIB \citep{franceschini2001} exceeds the
contribution of local {\it IRAS} galaxies by a factor of $\sim3$
\citep{soifer1991}.

In this paper we examine the colour-luminosity correlation and
luminosity function of {\it IRAS} galaxies, which together are an
important reference for constraining models of galaxy evolution with a
new generation of sub-mm surveys at shorter wavelengths. It has been
known for some time that more luminous {\it IRAS} galaxies exhibit
warmer dust temperatures \citep[e.g.][]{soifer1991}. This relationship
is important for a class of phenomenological models
\citep[e.g.][]{blain1993,guiderdoni1997,blain1999,chary2001,malkan2001,rowan-robinson2001,
  lagache2003, lagache2004} that have been used to predict the source
counts and redshift distributions at FIR wavelengths for the {\it
  Spitzer Space Telescope} \citep{werner2004} and {\it Akari}
\citep{matsuhara2006}, and at sub-mm wavelengths, for instruments such
as SCUBA \citep{holland1999}, MAMBO \citep{kreysa2002}, LABOCA
\citep{kreysa2003}, {\it Herschel}/SPIRE \citep{griffin2006}, SCUBA-2
\citep{holland2006}, AzTEC \citep{wilson2008}, and BLAST
\citep{pascale2008}. These models often use the shape of the CIB as an
integral constraint, since the total surface brightness of galaxies at
each wavelength cannot exceed the measured diffuse background.  These
authors apply evolution to local luminosity functions to obtain
estimates of the redshift-dependent luminosity function $\Phi(L,z)$.
In order to compare these models with observations, spectral energy
distribution (SED) templates are adopted to extrapolate observed flux
densities from the rest-frame luminosities. In a number of cases, it
has been beneficial to fit data over a range of wavelengths by
dividing the local luminosity function into several discrete
populations, each with different SED templates and separate
evolutionary forms
\citep[e.g.][]{blain1999,rowan-robinson2001,lagache2003,lagache2004}. Recently
\citet{wall2008} demonstrated direct evidence for the presence of at
least two significant populations in a sample of sub-mm luminous
sources in GOODS-N.  The luminosity-colour correlation can be useful
for such models as a method for assigning dust temperatures to SED
templates as a function of luminosity. For example,
\citet{lagache2003} use the observed {\it IRAS} colour-luminosity
distribution of \citet{soifer1991}, and \citet{lewis2005} use the FIR
colour-luminosity distribution of \citet[][henceforth
C03]{chapman2003} inferred from a model fit to {\it IRAS} data.

C03 calculate $\Phi(L,C)$, the galaxy volume density as a function of
total 3--1100\,\micron\ luminosity, $L_\mathrm{T}$, and the
60--100\,\micron\ colour, $C \equiv \log(S_{60}/S_{100})$. They
formulate $\Phi(L,C)$ as the product of a luminosity function, and the
distribution in $C$ as a function of $L$. The two functions are fit
independently, with the latter being constrained directly from the
observed distribution of the ratio of broad-band {\it IRAS} 60 and
100\,\micron\ fluxes.

In this work we provide a more accurate measurement of the joint
colour-luminosity distribution using a single maximum-likelihood
optimization to solve for all of the model parameters
simultaneously. Our methodology also differs from that of C03 in
several other key respects: (i) rather than calculating $C$ with
observed {\it IRAS} broad-band fluxes we use rest-frame monochromatic
60 and 100\,\micron\ flux densities derived from fitted SEDs; (ii) we
use narrower-bandwidth 42.5--122.5\,\micron\ FIR luminosities instead
of 3--1100\,\micron\ Total Infrared (TIR) luminosities to minimize the
dependence of the fitted distribution on the choice of SED templates;
(iii) a correction for redshift evolution in the {\it IRAS} galaxy
population is applied; and (iv) we account for a bias against the
detection of cooler sources caused by the 60\,\micron\ selection
criterion for the sample. Our galaxy sample, SED fitting procedure,
and methods for calculating luminosities and volumes are described in
Section~\ref{sec:sample}. The luminosity function and
colour-luminosity distribution are calculated in
Section~\ref{sec:lf}. We discuss the choice of luminosity variable,
and its consequences, in Section~\ref{sec:lumvar}.  Finally, we
compare the local colour-luminosity correlation with the observed
values for high-redshift sub-mm galaxies in
Section~\ref{sec:evolution} and test a simple evolutionary
model. Throughout this paper a standard cosmology is adopted with
$\Omega_M$\,=\,0.23, $\Omega_\Lambda$\,=\,0.77 and
$H_0$\,=\,74\,km\,s$^{-1}$\,Mpc\,$^{-1}$.

\section{Sample Preparation}
\label{sec:sample}

We use the same flux-limited $S_{60} > 1.2$\,Jy {\it IRAS} sample of
\citet{fisher1995} as in C03. Their catalogue covers most of the sky
and provides 60 and 100\,\micron\ flux-densities as well as
spectroscopic redshifts for each galaxy. The cool, high-luminosity
region of the observed colour-luminosity plane found to contain a
large number of spurious sources in C03 has also been excised. We use
this sample to first calculate a non-parametric (binned) FIR
luminosity function, and then fit it with simple parametric models.

\subsection{SEDs, Luminosities and Colours}
\label{sec:lum}

To calculate rest-frame luminosities and colours from observed {\it
  IRAS} 60 and 100\,\micron\ flux densities, we follow the method of
\citet{saunders1990}. A single temperature modified blackbody SED is
assumed for each source, $S(\nu) = A \nu^\beta
B_\nu(T_{\mathrm{obs}})$. The dust emissivity index is fixed at $\beta
= 1.5$ which is consistent with typical values measured for local
ULIRGs with sub-mm follow-up
\citep[e.g.][]{dunne2000,klaas2001,yang2007}. All of the subsequent
analysis in this paper has also been repeated using values $\beta=1.0$
and $2.0$, and the variation in the results is well within the quoted
uncertainties. The remaining two parameters, the amplitude $A$, and
the observed temperature $T_{\mathrm{obs}}$, are then uniquely
determined from the observed $S_{60}$ and $S_{100}$.  For this fit we
take into account the broad {\it IRAS} passbands
\citep{beichman1988}. Bolometric luminosities are calculated by
integrating the fitted SED directly --- the bolometric flux emitted in
the rest-frame, $S_{\mathrm{bol}}$ is simply the integral of the
observed SED across the red-shifted band, $S_{\mathrm{bol}} =
\int_{c/\lambda_l(1+z)}^{c/\lambda_u(1+z)} S(\nu) d\nu$, where
$\lambda_l$\,=\,122.5\,\micron\ and $\lambda_u$\,=\,42.5\,\micron\ for
FIR fluxes, and $\lambda_l$\,=\,1100\,\micron\ and
$\lambda_u$\,=\,3\,\micron\ for TIR fluxes. Similarly the colour $C$
is calculated from the logarithm of the ratio of monochromatic flux
densities emitted at 60 and 100\,\micron\ in the rest-frame by the
model SED.

\begin{figure} 
\epsfig{file=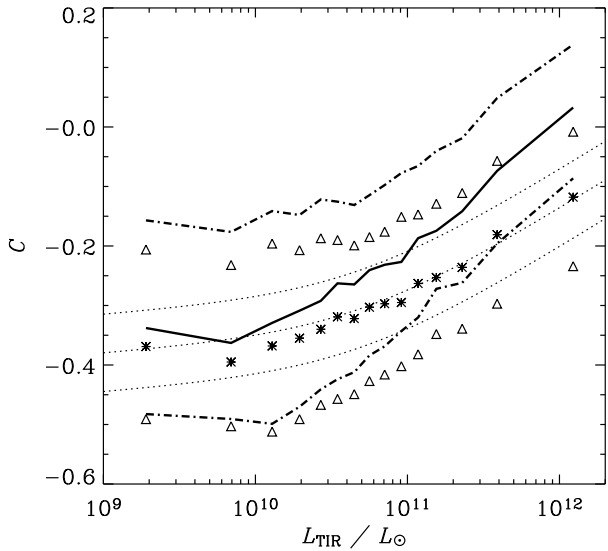,width=\linewidth}
\caption{Observed {\it IRAS} FIR colour distribution ($C\equiv
  \log(S_{60}/S_{100})$) as a function of 3--1100\,\micron\ TIR
  luminosity.  Stars and triangles show the mean and 68\% confidence
  intervals when $C$ is calculated from broad-band {\it IRAS} fluxes
  following the prescription in C03. The dotted lines show the mean
  and 1-$\sigma$ envelope of the fitted C03 colour-luminosity
  correlation for reference. The thick solid and dot-dashed lines show
  the mean and 68\% confidence intervals of the distribution when $C$
  is calculated with monochromatic flux densities emitted at 60 and
  100\,\micron\ in the rest-frame.}
\label{fig:obscol}
\end{figure}

Rather than a simple modified blackbody, C03 adopt the range of model
SED templates from \citet{dale2001}. This difference has a negligible
effect on the inferred FIR luminosities since there is very little
structure in the \citet{dale2001} SEDs at 42.5--122.5\,\micron\ that
is not characterized by the single temperature variable in our SED
model \citep[despite the correlation between luminosity and $\beta$
assumed in][]{dale2001}. For example, assuming temperatures ranging
from 30 to 50\,K (and $\beta=1.5$), the FIR luminosities inferred from
our modified blackbody templates compared with the \citet{dale2001}
SEDs with the same corresponding values of $C$ agree to within
$\sim5$\%. Since the difference is so small, and for the sake of
simplicity, we therefore proceed with the modified blackbody SED model
to measure the FIR colour-luminosity distribution. The TIR luminosity,
however, cannot be estimated from the modified blackbody model as
there is significant emission in the mid-infrared (MIR) spectrum
($\sim3$--60\,\micron) that is missed by the steep drop on the Wien
side \citep[e.g.][and discussion in
Section~\ref{sec:lumvar}]{blain2003}. Our modified blackbody SEDs in
this temperature range under-predict the TIR luminosities obtained
from the \citet{dale2001} SEDs by about 30\%. We only use their
templates to calculate TIR luminosities that are consistent with C03
for the discussion in this section (Figure~\ref{fig:obscol}), and
Section~\ref{sec:binned} (Figure~\ref{fig:lf_tir}).

Another fundamental difference between this work and C03 is the
definition of FIR colour. Whereas we choose to define $C$ in terms of
the ratio of rest-frame monochromatic flux densities, C03 use observed
broad-band {\it IRAS} fluxes. We believe our definition is more useful
as a general reference since no detailed knowledge of the {\it IRAS}
passbands is required in order to use our colour-luminosity
distribution. Furthermore, we find that the strength of the
colour-luminosity correlation is relatively diluted when using
broad-band fluxes. In Figure~\ref{fig:obscol} we show the distribution
of {\it IRAS} galaxy colours as a function of TIR luminosity derived
from fits of the \citet{dale2001} SED templates. The stars and
triangles correspond to the mean and 68\% confidence intervals using
observed broad-band {\it IRAS} fluxes, effectively re-producing the
top panel of Figure~1 in C03. The thick solid line and dot-dashed
lines show the mean and 68\% confidence interval of the colour
distribution using the ratio of monochromatic 60 and 100\,\micron\
flux densities emitted in the rest-frame. At luminosities $L \lsim
10^{10}$\,$L_\odot$ there is little difference in the shapes of the
distributions. At $L \gsim 10^{10}$\,$L_\odot$, however, the
colour-luminosity correlation is significantly steeper.

The dotted lines in Figure~\ref{fig:obscol} show the mean, and
1-$\sigma$ envelope of the C03 colour-luminosity correlation. We note
that although the mean of this fitted parametric distribution clearly
tracks the stars in the plot, the standard deviation of the
distribution, $\sigma_C = 0.065$, appears to have been
underestimated. We find that both the 68\% confidence intervals, and
the standard deviations of $C$, for each luminosity bin, are typically
closer to 0.13.

\subsection{Evolution in the Sample}
\label{sec:irasevol}

Since the most distant, luminous objects in {\it IRAS} samples exhibit
the effects of strong luminosity and/or density evolution
\citep[e.g.][]{saunders1990,kim1998,lawrence1999} we must account for
its effect in our measurement of the local luminosity function. Rather
than fitting for this evolutionary form ourselves, we instead apply an
explicit correction based on the luminosity evolutionary form fit by
\citet{saunders1990}: the luminosity of each galaxy is divided by a
factor $(1+z)^3$ corresponding to its redshift. We have chosen to
apply a luminosity, rather than a density evolution correction for
consistency with the discussion in Section~\ref{sec:evolution}.

\subsection{Accessible Volumes}
\label{sec:vol}

The $\sum(1/V_{\mathrm{max}})$ estimator \citep{schmidt1968}, with
accessible volumes $V_{\mathrm{max}}$ corresponding to the largest
redshift at which a galaxy would be detected given the survey flux
limit, used in C03, is appropriate for the monochromatic 60\,\micron\
luminosity function.  In this case a given object's luminosity,
$L_{60}$, is a function only of the observed flux density and
distance.  Therefore the maximum volume in which the given object can
be detected corresponds to the distance at which the observed
$S_{60}$, given its $L_{60}$, drops below the flux limit of the
sample. For the broad-band FIR luminosity function described here,
however, the relationship between $L_\mathrm{F}$ and $S_{60}$ is more
complicated and this simple method is invalid. There exists a bias
against the detection of cooler sources given the shape of the galaxy
SEDs \citep{saunders1990}. The wavelength of the peak FIR emission is
typically in the range 60--200\,\micron. The SEDs of warmer objects
peak closer to 60\,\micron, and colder objects at longer wavelengths,
so that in general for a fixed 60\,\micron\ flux density a colder
object must be {\em more FIR luminous} to be included in the sample.

\cite{saunders1990} derive the FIR luminosity function from their
60\,\micron\ flux-limited survey by selecting a sub-sample of objects
brighter than a FIR flux limit that corresponds to their 60\,\micron\
flux-limit {\em and} the coolest dust temperature that they observed,
23\,K \citep[see Section~6.5 in][]{saunders1990}.  However, this
selection reduces the size of their sample from $\sim3000$ objects, to
1004.  There is also an underlying assumption that there is no
significant population of sources with dust temperatures $T < 23$\,K.

In this work we use the entire sample, but calculate accessible
volumes using a modified formalism that accounts for the dependence of
$L_\mathrm{F}$ on the FIR colour. Given an observed temperature,
$T_\mathrm{obs}$, and redshift, $z$, the rest-frame temperature, $T$,
and total luminosity for an object are calculated. The accessible
volume corresponds to the maximum redshift at which an object with its
rest-frame luminosity {\it and} temperature would be detected in the
sample, or correspondingly the distance at which its observed flux
density in the {\it IRAS} 60\micron\ passband is 1.2\,Jy.

\section{The $\Phi(L,C)$ Distribution }
\label{sec:lf}

\subsection{Non-Parametric (binned) Estimate}
\label{sec:binned}

\begin{figure} 
\epsfig{file=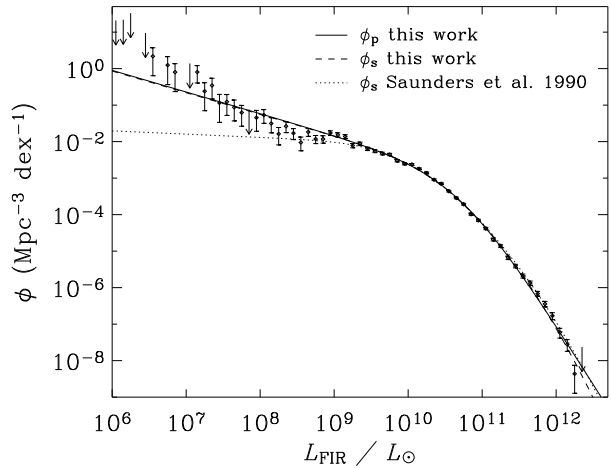,width=\linewidth}
\caption{The non-parametric 42.5--122.5\,\micron\ FIR luminosity
  function (symbols with 68\% Poisson error bars, and arrows show 95\%
  upper-limits for bins with $<$\,2 objects) and two parametric fits
  (Equation~\ref{eq:power_lumfunc} as a solid line, and
  Equation~\ref{eq:saunders_lumfunc} as a dashed line) derived from
  the $S_{60} > 1.2$Jy {\it IRAS} sample of \citet{fisher1995}.  The
  parametric FIR luminosity function of \citet{saunders1990} is shown
  for comparison as a dotted line.  }
\label{fig:lf_fir}
\end{figure}

With luminosities and accessible volumes for all of the objects in the
sample, we first calculate the non-parametric (binned)
colour-luminosity distribution, $\Phi_{\mathrm{b}}(L,C)$. Since the
accessible volume is now parameterized by both $L$ and $C$, the
modified $\sum(1/V_{\mathrm{max}})$ estimator is simply ,

\begin{equation}
  \Phi_{\mathrm{b}}(L,C) dL dC = \frac{4\pi}{\Omega_s}\sum_i \frac{1}{V_i},
\label{eq:lfbi}
\end{equation}

\noindent where $\Phi_{\mathrm{b}}(L,C) dL dC$ is the number of
sources in the area of the $L$--$C$ plane, and the sum runs over all
of the galaxies, $i$, with luminosity-evolution corrected luminosities
(Section~\ref{sec:irasevol}) and colours that land within the bin, and
with accessible volumes $V_i$.  The factor in front of the sum is the
fraction of the sky covered by the survey. The binned luminosity
function may be derived from this distribution by marginalizing over
$C$,

\begin{equation}
  \Phi_{\mathrm{b}}(L) dL = \sum_j \Phi_{\mathrm{b}}(L,C) dL dC,
\label{eq:lfbinned}
\end{equation}

\noindent where $j$ runs over all of the bins along the $C$ axis. We
define a second representation of the luminosity function using a
lower-case $\phi$,

\begin{equation}
  \phi(L) = \ln(10)L \Phi(L),
\end{equation}

\noindent which changes the units from Mpc$^{-3}$\,L$_\odot^{-1}$ to
the more typical Mpc$^{-3}$\,dex$^{-1}$, in order to assist comparison
with other work. This representation of our FIR luminosity function is
shown in Figure~\ref{fig:lf_fir}.

At luminosities $L \gsim 10^{9}\,L_{\odot}$ there is excellent
agreement between our luminosity function and the measurement of
\citet{saunders1990} (shown in Figure~\ref{fig:lf_fir} with a dotted
line). However, at fainter luminosities our luminosity function
includes many more objects. The reason for this discrepancy is
probably due to our choice of $\sum(1/V_{\mathrm{max}})$ estimator,
and the fact that over this luminosity range the sample is dominated
by an over-density of galaxies in the Local Supercluster. It is for
this reason that \citet{saunders1990} used an alternative estimator
that is insensitive to local density variations. Their method has the
potential to more accurately determine the {\it shape} of the
luminosity function, however at the expense of losing the absolute
normalization.  At luminosities $L \gsim 10^{9}\,L_{\odot}$ the
galaxies are typically sufficiently distant that this issue is no
longer important, and both estimators give consistent answers
\citep[see Section~8 of][]{lawrence1999}. In addition to this effect,
\citet{yun2001} suggest that some of the flattening at faint
luminosities in the \citet{saunders1990} luminosity function may be
due to sample incompleteness.

\begin{figure} 
\epsfig{file=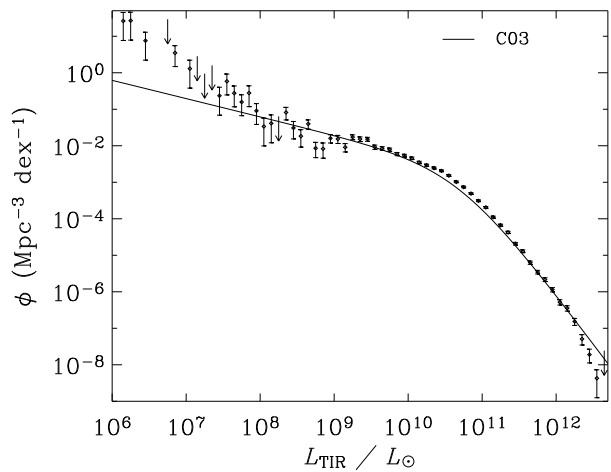,width=\linewidth}
\caption{The non-parametric 3--1000\,\micron\ TIR luminosity function
  (symbols with 68\% Poisson error bars, and arrows show 95\%
  upper-limits for bins with $<$\,2 objects) and the parametric fit of
  C03 (solid line). The discrepancies at $L > 10^9$\,$L_\odot$ are due
  to the corrections described in Sections~\ref{sec:irasevol} and
  \ref{sec:vol}.}
\label{fig:lf_tir}
\end{figure}

We also produce the non-parametric TIR luminosity function for
comparison with the parametric form of C03 in
Figure~\ref{fig:lf_tir}. For this calculation we have used the same
\citet{dale2001} SED templates as C03 to derive TIR luminosities. The
C03 model\footnote{We take the luminosity function to be
  $\sigma_C(2\pi)^{1/2}\Phi_1(L)$ from Section~3.2 in C03. The
  dimensionless pre-factor is needed since their colour distribution,
  $\Phi_2(C)$, is an un-normalized Gaussian with standard deviation
  $\sigma_C$. Also, we have assumed that the units for $\rho_*$ are
  Mpc$^{-3}$\,dex$^{-1}$ rather than Mpc$^{-3}$L$_\odot^{-1}$ as
  indicated.}  has a significantly different shape compared with our
binned representation at luminosities $L > 10^9$\,$L_\odot$, the range
over which the local over-density of galaxies is irrelevant. Their
model under-predicts the binned luminosity function by a factor
$\sim30$\% at $5\times10^{10}$\,$L_\odot$, and rises to over-predict
by a similar factor at $2\times10^{12}$\,$L_\odot$. We have determined
that this discrepancy can be explained entirely by the effects
described in Sections~\ref{sec:irasevol} and \ref{sec:vol}, since the
model and the bins are otherwise consistent without them. Applying
only the correction for evolution in the sample we find that the
number of objects in the brightest bins decreases -- explaining the
factor of 30\% at luminosities $>10^{12}$\,$L_\odot$. The reason for
this is that these objects are the most distant, and therefore exhibit
the strongest effects of redshift evolution. This correction has
almost no effect by $10^{11}$\,$L_\odot$. In contrast, applying the
correction for accessible volumes increases the numbers of objects in
bins at luminosities primarily $\lsim10^{11}$\,$L_\odot$. This
increase is caused by the fact that less luminous objects are cooler,
with correspondingly smaller volumes in which they could be detected
given the 60\,\micron\ flux limit. Together, these corrections
demonstrate that the luminosity function is in fact significantly
steeper than the result of C03 at luminosities
$\gsim10^{10}$\,$L_\odot$, the most important range for comparison
with results from new sub-mm surveys of distant star-forming galaxies.

\subsection{Maximum Likelihood Model Fits}

Next we fit simple parametric models to the data by maximizing the
likelihood of observing the sample. For the remainder of the paper we
consider only FIR luminosities to avoid dependence on assumptions
about the shape of the mid-infrared SED (Section~\ref{sec:lumvar}).
At a given position in the $L$--$C$ plane, the expected number of
sources from our sample to have landed in that bin, given a model for
$\Phi(L,C)$, is

\begin{equation}
\mu(L,C) = V_{\mathrm{max}}\frac{\Omega_s}{4\pi} \Phi(L,C) dL dC .
\end{equation}

\noindent These expectations are used to calculate the joint Poisson
likelihood of the data.

We express the model, $\Phi(L,C)$, as the product of the luminosity
function, $\Phi(L)$, and the conditional probability of a galaxy
having a color $C$ given the luminosity $L$, $p(C|L)$,

\begin{equation}
  \Phi(L,C) = \Phi(L) p(C|L).
\label{eq:factor}
\end{equation}

\subsubsection{Parametric forms of $\Phi(L)$}

For $\Phi(L)$, we consider two forms.  The first is the dual power-law
of C03,

\begin{equation}
\Phi_p(L) = \rho_* \left(\frac{L}{L_*}\right)^{(1-\alpha)}
\left(1 + \frac{L}{L_*}\right)^{-\beta},
\label{eq:power_lumfunc}
\end{equation}

\noindent where $L_*$ is the characteristic knee luminosity, $\rho_*$
is the number density normalization of the function at $L_*$
(Mpc$^{-3}$\,L$_\odot^{-1}$), and $\alpha$ and $\beta$ characterize
the power-laws at the faint ($L < L_*$) and bright ($L > L_*$) ends
respectively of the luminosity function.

The other form considered is the hybrid power-law/Gaussian form
preferred by \citet{saunders1990},

\begin{eqnarray}
  \nonumber
  \Phi_s(L) = \rho_* \left( \frac{L}{L_*} 
  \right)^{(1-\alpha)}
  \exp\left[ - \frac{1}{2\sigma^2} 
    \log^2_{10}\left(1 + \frac{L}{L_*}\right)\right] \times \\ 
    \frac{1}{\ln(10)L}, 
\label{eq:saunders_lumfunc}
\end{eqnarray}

Maximum likelihood solutions for both forms are shown in
Figure~\ref{fig:lf_fir}. At the faint end ($L_{\mathrm{FIR}} < L_*$)
both functions approach power-laws, and are therefore
indistinguishable. At the extreme bright-end they diverge; $\phi_s$
curves below $\phi_p$ at $L \sim 10^{12}\,L_\odot$, although both
forms lie mostly within the error bars of the non-parametric estimate.

To characterize the quality of the fits we calculate values of reduced
$\chi^2$ for luminosity bins that contain at least 10 objects,
approximately luminosities $5\times10^{8}$--$10^{12}$\,$L_\odot$ (with
this number of objects the Poisson error distribution is reasonably
approximated by a Gaussian). Over this range the power-law form
produces a value of reduced $\chi^2=2.2$, and the hybrid form
2.4. Given the similarity of these values, and the fact that each form
has the same number of parameters, we feel there is no compelling
evidence to favour one model over the other given the data. While the
choice has no impact on the subsequent discussion in this paper, we
note that the two forms rapidly diverge at luminosities
$>10^{12}$\,$L_\odot$, potentially the most important region of the
luminosity function for comparison with the results of sub-mm
surveys. For example, while at $10^{12}$\,$L_\odot$ the power-law only
exceeds the \citet{saunders1990} form by about 10\%, at
$10^{13}$\,$L_\odot$, it is nearly an order-of-magnitude
larger. Fitted parameters for both models are given in
Appendix~\ref{sec:a}, and they should only be considered valid to a
maximum luminosity of $\sim2\times10^{12}$\,$L_\odot$.

\subsubsection{Parametric form of  $p(C|L)$}

In C03 it was shown that the distribution in $C$ is approximately
Gaussian at a particular value of $L$. The precise functional form we
have adopted is

\begin{equation}  
p(C|L) = \frac{1}{\sigma_C\sqrt{2\pi}}\exp \left[ -\frac{1}{2} \times 
       \left( \frac{C-C_0}{\sigma_C} \right)^2 \right],
\label{eq:colour}
\end{equation}

\noindent with the mean colour at a given luminosity given
by\footnote{We note that in Section~3.2 of C03 the expression for
  $C_0$ is clearly meant to be the logarithm of the third equation in
  Section~3.1 -- the form we have adopted here.}

\begin{equation}
  C_0 = C_* - \delta \log_{10} \left( 1 + \frac{L'}{L}\right) + 
  \gamma \log_{10} \left( 1 + \frac{L}{L'}\right).
\label{eq:meancol}
\end{equation}

\noindent Note that unlike C03, the ``knee'' luminosity, $L'$, for
$p(C|L)$ is independent of the knee luminosity for the luminosity
function, $L_*$. In addition, the width of the colour distribution,
$\sigma_c$, is characterized by two different values: $\sigma_f$ and
$\sigma_b$ at the faint and bright ends respectively, with a smooth
transition at $L'$,

\begin{equation}
\sigma_c = \sigma_f(1-2^{-L'/L}) + \sigma_b(1-2^{-L/L'}). 
\label{eq:colsig}
\end{equation}

\begin{figure} 
\epsfig{file=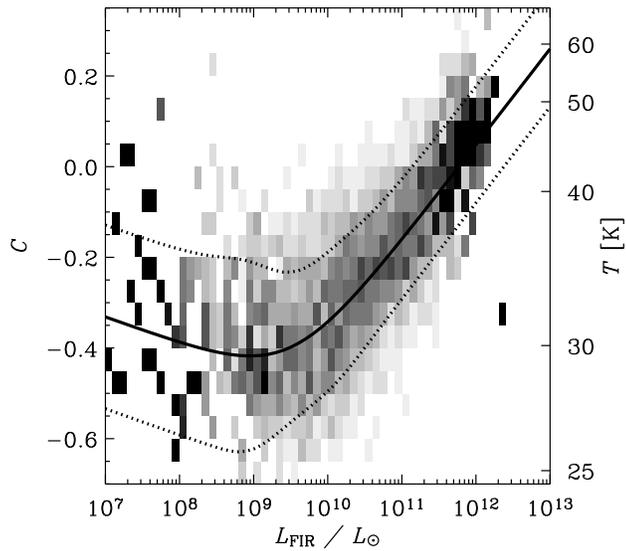,width=\linewidth}
\caption{Comparison between parametric (solid and dotted lines give
  the mean and 1-$\sigma$ envelope of Equation~\ref{eq:colour}
  respectively) and non-parametric estimates of $p(C|L)$ (greyscale
  shows $\Phi_{\mathrm{b}}(L,C)$ normalized along the $C$ axis).  The
  temperature axis is derived from $C$ assuming a dust emissivity
  index $\beta=1.5$.  At luminosities $L \lsim 10^{8}$ and $L \gsim
  10^{12}$ the sample does not contain enough galaxies to accurately
  constrain the shape, and $p(C,L)$ is simply extrapolated in the
  parametric model. }
\label{fig:p_lc}
\end{figure}

\noindent Figure~\ref{fig:p_lc} compares the mean and 1-$\sigma$
envelope of the parametric $p(C|L)$ with a non-parametric estimate
created by factoring the smooth model $\Phi(L)$ from the binned
$\Phi_{\mathrm{b}}(L,C)$.

\begin{figure} 
\epsfig{file=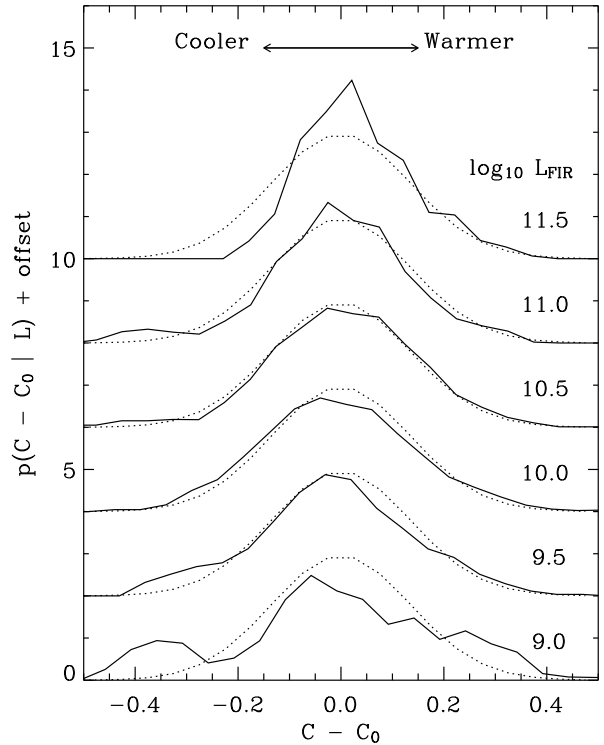,width=\linewidth}

\caption{The distribution of FIR colours, $C$, about the mean, $C_0$
  (Equation~\ref{eq:meancol}), at a range of luminosities. The solid
  lines are normalized slices of the measured (non-parametric)
  $p(C|L)$ ($\Phi_{\mathrm{b}}(L,C)/\Phi_{\mathrm{b}}(L)$ -- See
  Equations~\ref{eq:lfbi} and \ref{eq:factor}) evaluated at $C-C_0$,
  and numbers indicate $\log_{10}(L_{\mathrm{FIR}})$. The parametric
  model (Equation~\ref{eq:colour}) is shown as a dotted line.}
\label{fig:coldist}
\end{figure}

It is argued in C03 that the width of the distribution in $C$ is
constant as a function of luminosity. The top panel of Figure~2 in C03
demonstrates a constant width in $S_{60}/S_{100}$ with a logarithmic
axis, in contrast with the bottom panel in which the width of the
distribution is shown to broaden at greater luminosities when plotted
with a linear axis. This behaviour motivates the definition
$C\equiv\log(S_{60}/S_{100})$. However, this plot appears to be at
odds with the top panel of Figure~4 in C03 which exhibits a systematic
broadening at higher luminosities. Such a trend does not appear to be
present in our measurement of $p(C|L)$. For clarity, we compare slices
of the parametric estimate of $p(C|L)$ at several fixed luminosities
with the binned estimate in Figure~\ref{fig:coldist}. The good
agreement between these two estimates, both in terms of scatter and
systematic variations, indicates that Equation~\ref{eq:colour}
adequately describes the shape of $p(C|L)$. We find that the width of
the distribution narrows with increasing brightness to $\sigma_b=0.13$
from $\sigma_f=0.2$ at a transition luminosity of
$\sim3.5\times10^9$\,$L_\odot$ (Appendix~\ref{sec:a}). The broadening
shown in C03 at greater luminosities does not appear to be caused by
their choice of SEDs or choice of broad-band over monochromatic
colours. The most likely explanation is an artifact of the C03 griding
scheme. They use variable-width luminosity bins which contain equal
numbers of objects, in contrast to our method which uses
equally-spaced logarithmic bins. The wide, sparsely-populated
high-luminosity bins may simply dilute the colour-luminosity
correlation.  We note that our fitted value for $\sigma_b$ appears to
be consistent with the width of the distribution in the lowest,
narrowest, luminosity bins in the top panel of Figure~4 from C03,
(although they claim a smaller standard deviation of $0.065$; see
Figure~\ref{fig:obscol} and the discussion at the end of
Section~\ref{sec:lum} in this paper).

\subsubsection{Parameter Uncertainties}
\label{sec:uncertainties}

We characterize the uncertainties in the ten parameters of
$\Phi(L,C)$, for both parametric forms of $\Phi(L)$, using a bootstrap
Monte Carlo technique.  First, 100 realizations of the 1.2\,Jy survey
are created from the actual survey data by randomly sampling sources
from the catalogue with replacement \citep[see Section~6.6
of][]{wall2003}. We then fit the model to each simulated sample. From
these 100 fits we calculate the sample variances, and covariances
between all pairs of parameters to estimate the full parameter
covariance matrix. The maximum likelihood values of each parameter,
their standard deviations, and the Pearson correlation matrices are
given in Appendix~\ref{sec:a}. Note that the parameters for $p(C|L)$
are largely independent of the parametric form chosen for $\Phi(L)$.

\section{Discussion}

\subsection{Choice of Bolometric Luminosity Variable}
\label{sec:lumvar}

In this work we have chosen to use the 42.5--122.5\,\micron\ FIR
luminosity instead of the wider-bandwidth 3--1100\,\micron\ total
infrared luminosity, $L_\mathrm{T}$, as in C03. An argument for using
the latter is that it includes a significant fraction of the total
power emitted by a galaxy that is missed at shorter wavelengths ($<
40$\,\micron) --- an effect which becomes increasingly important at
high luminosities given the positive $L$--$C$ correlation. By using
$L_{\mathrm{FIR}}$, the ``clipping'' of shorter-wavelength light
obscures the physical interpretation of the correlation between the
two variables.

\begin{figure} 
\epsfig{file=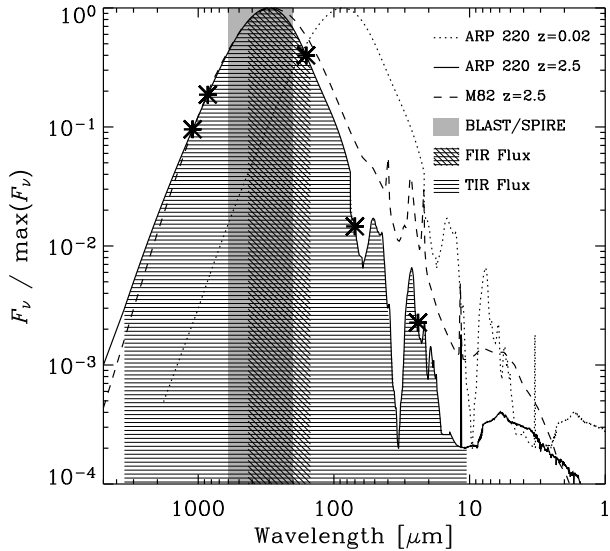,width=\linewidth}
\caption{The peak-normalized spectral energy distribution of Arp~220
  (dotted line), and shifted to $z=2.5$ (solid line). The complete
  observed BLAST and SPIRE band (200--600\,\micron, including the 30\%
  finite bandpasses for each channel) is indicated by the solid grey
  shaded rectangle.  The rest-frame 3--1100\,\micron\ total infrared
  (TIR) flux, and 42.5--122.5\,\micron\ FIR flux correspond to the
  integrals of the horizontal-line filled, and cross-hatch filled
  regions respectively, for the $z=2.5$ SED.  Clearly for objects at
  this redshift, the BLAST and SPIRE filters more closely match the
  rest-frame FIR flux than the TIR flux. The peak-normalized M82 SED
  redshifted to $z=2.5$ (shown as a dashed line) illustrates the
  relatively large variations in the rest-frame mid-infrared spectrum
  ($\sim3$--60\,\micron\ observed wavelength).  Symbols indicate the
  wavelengths accessible from ground-based sub-mm surveys (e.g. SCUBA
  at 850\,\micron, and BOLOCAM/AzTEC/MAMBO at 1.1\,mm), and existing
  space-based FIR data (e.g. {\it Spitzer} at 24, 70 and
  160\,\micron), demonstrating their inability to accurately constrain
  the bolometric FIR luminosity for these galaxies.}
\label{fig:sed_bands}
\end{figure}

Our primary goal, however, is to assist with the development of a
model for the luminosity function of SMGs discovered in BLAST
\citep{pascale2008} and future {\em Herschel}/SPIRE
\citep{griffin2006} 250--500\,\micron\ surveys, as well as any other
surveys at similar wavelengths. It is now generally accepted that the
redshift distribution for the bulk of SMGs discovered in 850\,\micron\
SCUBA surveys peaks at redshifts $\sim2.5$. For example, the
radio-detected spectroscopic sample of \citet{chapman2003,chapman2005}
finds a median redshift of 2.2 with an interquartile range
$z=1.7$--2.8, in general agreement with several other studies using
radio--FIR or radio and 24\,\micron\ guided photometric redshift
estimates \citep[e.g.][]{aretxaga2003,aretxaga2007,pope2006}.  Since
the negative $K$-correction produces a nearly un-biased detection
efficiency at 850\,\micron\ for a typical ULIRG SED at $z\sim1$--8
\citep[e.g.][]{blain2002}, the observed redshift distribution is a
reasonable proxy for the total dust-obscured star-formation rate
history of massive galaxies. If SMGs have thermal SEDs similar to the
ULIRGs that populate the bright end of the {\em IRAS} luminosity
function presented here, their SEDs peak at wavelengths
$\sim60$--200\,\micron\ in the rest-frame, and they are redshifted
into the 200--600\micron\ BLAST and SPIRE bandpasses near the peak of
their redshift distribution.

As an example, Figure~\ref{fig:sed_bands} shows the SED of the ULIRG
Arp~220 at a redshift $z=2.5$ compared to the BLAST and SPIRE bandpass
region, and the integrated FIR and TIR fluxes. Clearly the BLAST and
SPIRE integrated fluxes more closely matches the rest-frame FIR than
TIR flux.

The peak-normalized SED of the starburst galaxy M82 redshifted to
$z=2.5$ is shown for comparison as a dashed line to illustrate the
relatively large scatter at mid-infrared (MIR) wavelengths compared to
the smooth thermal SED at longer wavelengths.  The rest-frame MIR SEDs
of these example galaxies exhibit prominent polycyclic aromatic
hydrocarbon (PAH) absorption and emission features \citep[the
$\sim100$--10\,\micron\ range of the observed SED; measurements for
actual SCUBA-selected SMGs are given in][]{pope2008}. For these two
examples there is a difference of $\sim30$\% in the contribution of
the mid-infrared emission to the TIR luminosity.  For these reasons,
evolutionary models based on the FIR luminosity function are less
dependent on assumptions about the intrinsic near-IR to
millimetre-wavelength SEDs of high-redshift galaxies than the TIR
luminosity function, enabling cleaner comparison with new and future
data from deep sub-mm cosmological surveys.  We emphasize the fact
that one is free to adopt any template library to infer flux densities
at other wavelengths for objects drawn from our $\Phi(L,C)$
distribution provided that they have roughly thermal FIR spectra, and
span the relevant range of rest-frame FIR colours $-0.65 \lsim C \lsim
0.25$ \citep[see for example the comparison in Section~\ref{sec:lum}
between our modified blackbody SED model and the library
of][]{dale2001}.

\subsection{Evidence for an Evolving Colour-Luminosity Correlation}
\label{sec:evolution}

\begin{figure} 
\epsfig{file=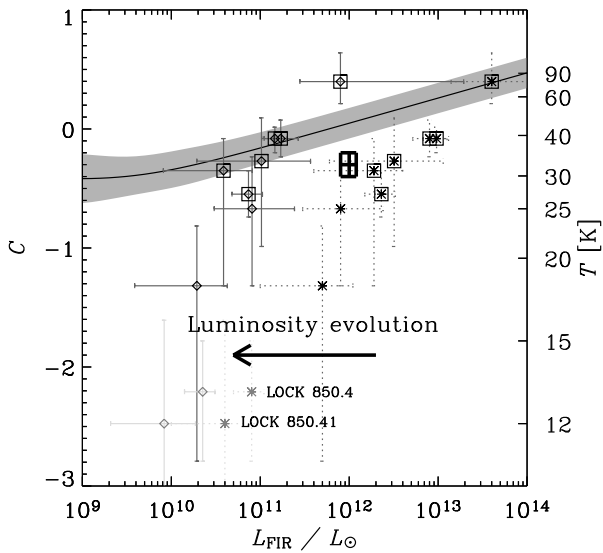,width=\linewidth}
\caption{Comparison of local $p(C|L)$ (shaded region is the 68\%
  confidence interval, solid line is $C_0$ from
  Equation~\ref{eq:meancol}) with data from \citet{coppin2008}. The
  temperature axis is derived from $C$ assuming a dust emissivity
  index $\beta=1.5$.  Stars with dotted 1-$\sigma$ error bars indicate
  the 10 SMGs with spectroscopic redshifts and temperatures derived
  from observed 350\,\micron/850\,\micron\ colours. Squares have been
  drawn around the symbols for the 6 objects at $z>2$.  High-redshift
  ultra-luminous galaxies appear systematically cooler than those in
  the local universe. Under the assumption of pure luminosity
  evolution of the form $(1+z)^3$, the SMGs have been projected into
  the local colour-luminosity distribution by shifting them along the
  luminosity axis (diamonds with solid error bars). The model is a
  plausible fit to the data except for LOCK~850.4 and LOCK~850.41
  which appear much cooler (shown as lighter symbols).  These objects
  may have ambiguous or incorrect optical/IR counterpart
  identifications (discussed in Section~\ref{sec:evolution}). Future
  BLAST and SPIRE surveys will constrain FIR luminosities of
  $\sim10^{12}$\,$L_\odot$ galaxies to $\sim20$\%, with uncertainties
  in $C$ of $\sim0.1$ (several Kelvin). A representative measurement
  in the $L$--$C$ plane is given by the thick cross-filled square. }
\label{fig:smg_col}
\end{figure}

Evolution in the FIR colour-luminosity distribution is difficult to
probe from {\it IRAS} catalogues given their relatively low
redshifts. For example, the median galaxy redshift in the
\citet{fisher1995} sample is $z=0.019$, and the most distant object is
at $z=0.326$. Despite this, it has been possible to place weak
constraints on the evolutionary form of the FIR luminosity
function. \citet{saunders1990} found that the most distant galaxies
could undergo extremes of pure density evolution of the form
$(1+z)^{7\pm2}$, or pure luminosity evolution of the form
$(1+z)^{3\pm1}$ (we use this latter form explicitly to correct our
sample, see Section~\ref{sec:lum}). Since the most distant {\it IRAS}
galaxies are also the brightest in the sample (luminosities $>L_*$) it
is not possible to determine which form (or combination) is the more
relevant.

At higher redshifts, the best constraints on the evolution of
ultra-luminous infrared galaxies come from SCUBA surveys at
850\,\micron. It can be shown, for example, that continued luminosity
evolution in the bright-end of the local FIR luminosity function of
the same form as \citet{saunders1990} to redshifts $\sim2$--3 can be
used to fit mm--submm source counts, assuming ULIRG-like SEDs to
extrapolate observed flux densities from rest-frame luminosities
\citep[e.g][]{scott2002,lagache2003}. A more direct test of evolution
for a small sample of SMGs with known redshifts was recently performed
by \citet{wall2008} finding similar results.  Unfortunately, until
recently, shorter-wavelength data that would help to constrain the
dust temperatures of objects in these samples is generally
unavailable, and it is therefore not possible to search directly for
{\it colour} evolution. Measuring dust temperatures for large samples
($\gg 1000$) of SMGs is one of the primary science goals of BLAST and
{\it Herschel}/SPIRE surveys --- with the caveat that redshifts must
first be determined independently for at least a subset of these new
objects since there is a potential degeneracy between the apparent
observed temperature and redshift \citep[e.g.][]{blain2003}.

Recent SHARC-II 350\,\micron\ observations of SMGs
\citep[e.g.][]{kovacs2006,coppin2008}, however, have enabled improved
estimates of dust temperatures for smaller samples (several tens of
galaxies).  In Figure~\ref{fig:smg_col} ten objects (triangles) with
constrained dust temperatures and spectroscopic redshifts (median
$z=2.1$) from \citet{coppin2008} are compared with our local
measurement of $p(C|L)$. As noted in \citet{kovacs2006} and
\citet{coppin2008}, the rest-frame temperatures for such luminous
galaxies (median $L_{\mathrm{FIR}}=2.3\times10^{12}$\,$L_\odot$) are
much lower than objects in the local Universe.  With the correlation
between luminosity and FIR colour in-hand, we now ask the question:
can pure luminosity evolution of the form $(1+z)^3$ account for the
apparently cooler temperatures of SMGs at high-redshift? Explicitly,
we express the redshift evolution of the FIR colour-luminosity
distribution, $\Phi(L,C,z)$, as a simple function of the local
distribution,

\begin{equation}
\Phi(L,C,z) = \Phi\left(\frac{L}{(1+z)^3},C\right).
\end{equation}

\noindent For comparison, we project each observed SMG into the local
$p(C|L)$ distribution by dividing their luminosities by $(1+z)^3$,
shifting the objects to the left in Figure~\ref{fig:smg_col}
(diamonds). Given the uncertainties, the 8 warmer objects (top of the
plot) are roughly consistent with the local distribution once we apply
this transformation. Ignoring the redshift uncertainties and adding
the measured colour uncertainties, $\sigma_{\mathrm{data}}$, in
quadrature with the intrinsic colour width in the source population,
$\sigma_c$ (Equation~\ref{eq:colsig}, at the evolution-corrected
luminosity of the galaxy -- the shaded region of the figure), we
calculate residuals, $R$, between the model, $C_0$ (solid line through
the centre of the shaded region) and the measured colours,
$C_{\mathrm{data}}$:

\begin{equation}
  R = \frac{C_{\mathrm{data}} - C_0}{\sqrt{\sigma_{\mathrm{data}}^2 + \sigma_c^2}}.
\end{equation}

\noindent We note that for five of these objects $R<1$, and the
remaining three are in the range $R=1$--1.5; the approximate
expectation for uncorrelated Gaussian uncertainties. This result is
contrary to the conclusion of C03 who found no need to invoke redshift
evolution for $p(C|L)$ when comparing with lower-redshift ($z<1$)
samples.

The two most significant outliers, LOCK~850.4 ($C$\,=\,-2.2,
$z$\,=\,0.526) and LOCK~850.41 ($C$\,=\,-2.47, $z$\,=\,0.689), also
the two coolest and least luminous objects, deserve further
explanation. These two galaxies are also the only objects with
redshifts $z<1$. Since their observed sub-mm colours are otherwise
similar to the other galaxies, these low redshifts also imply lower
rest-frame dust temperatures (lower values of $C$). There is a
possibility that the true optical counterpart (and hence redshift) for
LOCK~850.4 is at $z$\,=\,1.482, rather than 0.526 as adopted by
\citet{coppin2008}. Both potential counterparts were proposed in
\citet{ivison2005}. Adopting the higher redshift object as the
counterpart, the inferred FIR luminosity increases from
$8\times10^{10}$\,$L_\odot$ to $10^{12}$\,$L_\odot$ and the rest-frame
dust temperature from 13\,K to 21\,K, or $C$\,=\,-1, at which point it
would appear to have similar dust properties to the other galaxies in
the sample.  There is a similar possibility of a mis-identification
for LOCK~850.41. Two counterparts are suggested in \citet{ivison2005},
although they were only able to obtain the spectroscopic redshift
indicated above for one of them. The other counterpart has an optical
photometric redshift estimate of $z$\,=\,2.2$\pm$0.2 from
\citet{dye2008}. A similar value of $z$\,=\,$2.1\pm^{1.4}_{0.6}$ based
on its FIR colours is proposed in \citet{aretxaga2007}. Adopting a
redshift of $z$\,=\,2.2, the luminosity for LOCK~850.41 increases from
$4\times10^{10}$\,$L_\odot$ to $5\times10^{11}$\,$L_\odot$ and the
rest-frame dust temperature from 12\,K to 22\,K, or $C$\,=\,-0.9, also
closer to the distribution for the other objects in the
sample. However, if the lower-redshift candidates for these two
galaxies are correct, and a population of galaxies with temperatures
$T \lsim 15$\,K do exist in abundance at redshifts $z<1$, it is
possible that they were completely missed in {\it IRAS} surveys, and
will appear in the wide-area SCUBA-2 and BLAST and SPIRE surveys.

Finally, we note that the higher-redshift objects generally fall the
closest to the colour-luminosity distribution. To emphasize this fact
we draw squares around the 6 objects in this small sample at redshifts
$z>2$. Naively this fact is slightly surprising, since one might
suppose that the nearer objects are in fact more similar to the sample
used to constrain the local distribution. However, an additional
consideration is the selection function for this SCUBA sample. While
the negative $K$-correction produces approximately the same observed
850\,\micron\ flux density at redshifts $z\sim1$--8 for a fixed FIR
luminosity and temperature, there is also a bias towards the detection
of cooler objects at a fixed redshift and flux density since such
objects are less luminous, and hence more abundant in the rest-frame
\citep[e.g.][]{eales1999,blain2002,chapman2003}. For the
lower-redshift objects, at which point the negative $K$-correction is
diminished, the luminosities are also fainter for a given flux
density, and this bias could be increased due to the broadening in the
colour-luminosity correlation that we have measured at lower
luminosities.

This comparison is by no means an exhaustive study of evolution in the
FIR colour-luminosity distribution. It is our goal to extend this
investigation to much larger samples of SMGs with FIR colour
information in new BLAST extra-galactic surveys (Devlin et al. in
prep.), and future {\it Herschel}/SPIRE surveys. Understanding the
details of this evolution is intimately related to the star-formation
rate history of massive galaxies, since the rest-frame FIR luminosity
is the key observed quantity in SMGs from which star-formation rates
are derived
\citep[e.g.][]{kennicutt1998,hughes1998,scott2002,pope2006}.

\section{Conclusions}

We have measured the local FIR galaxy colour-luminosity distribution
based on the 60 and 100\,\micron\ flux densities and redshifts from
the \citet{fisher1995} all-sky {\it IRAS} sample.  This distribution
is an important reference for forthcoming BLAST and {\it
  Herschel}/SPIRE extra-galactic surveys at 250, 350, and
500\,\micron\ that will detect thousands of SMGs at redshifts
$z>1$. Since the space density of SMGs appears to peak at redshifts
$z\sim2.5$, defining the epoch at which most of the stars in
present-day massive galaxies formed, the BLAST and SPIRE bandpasses
will sample the same region of the rest-frame SEDs for these objects
as {\it IRAS} at 60 and 100\,\micron\ for local samples. Our
measurement is therefore the primary present-day boundary condition
with which any evolutionary model for the luminosity and
dust-temperature distribution of SMGs must be compared. This applies
to current BLAST and future SPIRE surveys, as well as
higher-resolution ground-based observations with SCUBA-2 for example.

Our method accounts for a temperature bias in the underlying
60\,$\mu$m flux-limited sample, as well as luminosity evolution. These
corrections indicate that the bright-end of the luminosity function is
significantly steeper than an earlier calculation by C03 which
neglected them.  We have verified that our distribution is consistent
with the FIR luminosity function of \citet{saunders1990} by
marginalizing over colour.  We fit a parametric model to the data
consisting of the product of the luminosity function, $\Phi(L)$, with
the conditional colour probability, $p(C|L)$, where $C\equiv
\log(S_{60}/S_{100})$. We fit $p(C|L)$ using a normal distribution for
$C$ as a function of $L$, with a mean colour given by a broken
logarithmic function also of $L$. This is characterized by the knee
luminosity for the break, $L'$, which is independent of the luminosity
function knee, $L_*$, and the width of the distribution, by two
different standard deviations $\sigma_b$ and $\sigma_f$, above and
below the knee $L'$.

We have checked directly for evolution in the colour-luminosity
correlation using observations of high-redshift SMGs ($z>1$) with
temperatures constrained by SCUBA 850\,\micron\ and SHARC-II
350\,\micron\ photometry from \citet{coppin2008}. These high-$z$
ultra-luminous objects appear much cooler than local galaxies of
comparable luminosities, and there is preliminary evidence that pure
luminosity evolution in the local colour-luminosity distribution of
the form $(1+z)^3$ is consistent with uncertainties in their measured
redshifts and colours. This result is contrary to C03 who find no
evidence for a change in the relationship between luminosity and
colour in low-redshift ($z<1$) samples.

\section{Acknowledgements}

We thank Manolis Plionis, Enrique Gazta\~naga, Min Yun and Douglas
Scott for valuable discussions. We also thank the anonymous referee
for their helpful comments. EC conducted a portion of this research
with the support of an NSERC Postgraduate Scholarship.  IA and DHH
acknowledge partial support from Conacyt grants 50786 and 60878.

\bibliographystyle{mn2e}
\bibliography{mn-jour,refs}

\appendix

\section{Model parameters and uncertainties}
\label{sec:a}

Maximum likelihood parameters and 1$\sigma$ uncertainties for the
local FIR colour-luminosity distribution are given in
Tables~\ref{tab:plawmean}1 and \ref{tab:saundersmean}3. Pearson
correlation coefficients for the uncertainties are given in
Tables~\ref{tab:plawerr}2 and \ref{tab:saunderserr}4.

\begin{table}
  \label{tab:plawmean}
  \caption{Maximum likelihood parameter values, and 1-$\sigma$ uncertainties for $\Phi(L,C)$ (Equation~\ref{eq:factor}) using the C03
    dual power-law form of the luminosity function (Equation~\ref{eq:power_lumfunc}, and Equations~\ref{eq:colour}--\ref{eq:colsig} for $p(C|L)$).}
  \centering
\begin{tabular}{cc}
  Parameter  & Value \\
  \hline
  $\rho_*$   & (1.22 $\pm$ $0.24) \times 10^{-14}$ Mpc$^{-3}$\,L$_\odot^{-1}$ \\
  $\alpha$   & 2.59  $\pm$ 0.03 \\
  $L_*$      & (5.14 $\pm$ $0.39) \times 10^{10}$ L$_\odot$ \\
  $\beta$    & 2.65  $\pm$ 0.05 \\
  $\sigma_b$ & 0.128 $\pm$ 0.003 \\
  $\sigma_f$ & 0.20  $\pm$ 0.01 \\
  $C_*$      & -0.48 $\pm$ 0.02 \\
  $\delta$   & -0.06 $\pm$ 0.02 \\
  $\gamma$   & 0.21  $\pm$ 0.01 \\
  $L'$       & (3.2 $\pm$ $1.7) \times 10^{9}$ L$_\odot$ \\ \hline
\end{tabular}
\end{table}

\begin{table*}
  \label{tab:plawerr}
  \caption{Parameter Pearson correlation matrix for $\Phi(L,C)$ using the C03
    dual power-law form of the luminosity function (see Table~\ref{tab:plawmean}1).}
\centering
\begin{tabular}{c|cccccccccc}
           & $\rho_*$ & $\alpha$ & $L_*$ & $\beta$ & $\sigma_b$ & $\sigma_f$ & $C_*$ & $\delta$ & $\gamma$ & $L'$ \\ \hline
$\rho_*$   &    1.00 & -0.88 & -0.93 & -0.24 &  0.03 &  0.03 & -0.02 &  0.04 &  0.06 &  0.05 \\
$\alpha$   &         &  1.00 &  0.73 & -0.13 &  0.05 &  0.00 & -0.04 & -0.06 & -0.18 & -0.13 \\
$L_*$      &         &       &  1.00 &  0.49 & -0.10 & -0.02 &  0.08 & -0.00 &  0.04 &  0.03 \\
$\beta$    &         &       &       &  1.00 & -0.23 &  0.03 &  0.17 &  0.10 &  0.31 &  0.22 \\
$\sigma_b$ &         &       &       &       &  1.00 &  0.09 & -0.77 & -0.74 & -0.71 & -0.76 \\
$\sigma_f$ &         &       &       &       &       &  1.00 & -0.45 & -0.29 & -0.39 & -0.47 \\
$C_*$      &         &       &       &       &       &       &  1.00 &  0.92 &  0.75 &  0.94 \\
$\delta$   &         &       &       &       &       &       &       &  1.00 &  0.67 &  0.83 \\
$\gamma$   &         &       &       &       &       &       &       &       &  1.00 &  0.89 \\
$L'$       &         &       &       &       &       &       &       &       &       &  1.00 \\
\end{tabular}
\end{table*}

\begin{table}
  \label{tab:saundersmean}
  \caption{Maximum likelihood parameter values, and 1-$\sigma$ uncertainties for $\Phi(L,C)$ (Equation~\ref{eq:factor}) using the
    \citet{saunders1990} form of the luminosity function (Equation~\ref{eq:saunders_lumfunc}, and Equations~\ref{eq:colour}--\ref{eq:colsig} for $p(C|L)$).}
\centering
\begin{tabular}{cc}
  Parameter  & Value \\
  \hline
  $\rho_*$   & (6.29 $\pm$ $0.64) \times 10^{-3}$ Mpc$^{-3}$\,dex$^{-1}$ \\
  $\alpha$   & 1.59  $\pm$ 0.03 \\
  $L_*$      & (3.99 $\pm$ $0.53) \times 10^{9}$ L$_\odot$ \\
  $\sigma$   & 0.60  $\pm$ 0.01 \\
  $\sigma_b$ & 0.127 $\pm$ 0.004 \\
  $\sigma_f$ & 0.20  $\pm$ 0.01 \\
  $C_*$      & -0.47 $\pm$ 0.03 \\
  $\delta$   & -0.05 $\pm$ 0.02 \\
  $\gamma$   & 0.22  $\pm$ 0.01 \\
  $L'$       & (3.8  $\pm$ $2.5) \times 10^{9}$  L$_\odot$ \\ \hline
\end{tabular}
\end{table}

\begin{table*}
  \label{tab:saunderserr}
  \caption{Parameter Pearson correlation matrix for $\Phi(L,C)$ using the
    \citet{saunders1990} luminosity function (see Table~\ref{tab:saundersmean}2).}
  \centering
\begin{tabular}{c|cccccccccc}
           & $\rho_*$ & $\alpha$ & $L_*$ & $\sigma$& $\sigma_b$ & $\sigma_f$ & $C_*$ & $\delta$ & $\gamma$ & $L'$ \\ \hline
$\rho_*$   &  1.00 & -0.73 & -0.98 &  0.83 &  0.27 &  0.33 & -0.34 & -0.33 & -0.31 & -0.38 \\
$\alpha$   &       &  1.00 &  0.75 & -0.34 &  0.09 & -0.11 &  0.01 &  0.07 & -0.09 &  0.02 \\
$L_*$      &       &       &  1.00 & -0.85 & -0.23 & -0.30 &  0.29 &  0.29 &  0.29 &  0.35 \\
$\sigma$   &       &       &       &  1.00 &  0.33 &  0.31 & -0.33 & -0.29 & -0.40 & -0.41 \\
$\sigma_b$ &       &       &       &       &  1.00 &  0.41 & -0.78 & -0.67 & -0.83 & -0.77 \\
$\sigma_f$ &       &       &       &       &       &  1.00 & -0.74 & -0.67 & -0.55 & -0.67 \\
$C_*$      &       &       &       &       &       &       &  1.00 &  0.92 &  0.83 &  0.93 \\
$\delta$   &       &       &       &       &       &       &       &  1.00 &  0.73 &  0.81 \\
$\gamma$   &       &       &       &       &       &       &       &       &  1.00 &  0.91 \\
$L'$       &       &       &       &       &       &       &       &       &       &  1.00 \\
\end{tabular}
\end{table*}

\end{document}